\documentclass[aps,prl,twocolumn,showpacs,10pt,superscriptaddress,floatfix,longbibliography]{revtex4-1}

\usepackage{newtxtext,newtxmath}
\usepackage{graphicx}
\usepackage{subfigure}
\usepackage{amsmath}
\usepackage{amsfonts}
\usepackage{amssymb}
\usepackage{mathrsfs}
\usepackage{bbm}
\usepackage{subfigure}
\usepackage{xcolor}
\usepackage[colorlinks=true]{hyperref}
\usepackage{ulem}
\usepackage[T1]{fontenc}

\begin{document}

\title{Berry curvature effects on quasiparticle dynamics in superconductors}

\author{Zhi Wang}
\altaffiliation{These authors contributed equally to this work.}
\affiliation{School of Physics, Sun Yat-sen University, Guangzhou 510275, China}

\author{Liang Dong}
\altaffiliation{These authors contributed equally to this work.}
\affiliation{Department of Physics, The University of Texas at Austin, Austin, Texas 78712, USA}

\author{Cong Xiao}
\email{Corresponding author: congxiao@utexas.edu}
\affiliation{Department of Physics, The University of Texas at Austin, Austin, Texas 78712, USA}

\author{Qian Niu}
\affiliation{Department of Physics, The University of Texas at Austin, Austin, Texas 78712, USA}

\begin{abstract}
We construct a theory for the semiclassical dynamics of
superconducting quasiparticles by following their wave-packet motion and reveal rich contents of Berry curvature effects in the phase-space spanned by position and momentum. These Berry curvatures are traced back to the
characteristics of superconductivity, including the nontrivial momentum-space
geometry of superconducting pairing, the real-space supercurrent, and the charge dipole of quasiparticles. The Berry-curvature effects strongly influence the spectroscopic and transport properties of superconductors, such as the local density of states and the thermal Hall conductivity.
As a model illustration, we apply the theory to study the twisted
bilayer graphene with a $d_{x^{2}+y^{2}}+id_{xy}$ superconducting gap function, and demonstrate Berry-curvature induced effects.
\end{abstract}
\maketitle


{\it {\color{blue} Introduction.}}--
The Chern number of Bogoliubov-de Gennes band structure has commonly been used to characterize the topology of exotic superconductors \cite{qi2011RMP,Black_Schaffer2014,Sato2017,liu2013,yang2018d+id}, while much less attention has been given to the physical effect of the momentum space Berry curvature
which makes up the Chern number \cite{cvetkovic2015,liang2017wavepacket}. In the presence of inhomogeneity due to
external fields or a supercurrent, we may also expect to find other components
of the Berry curvature in the phase space, such as those in the real space as
well as in the cross planes of position and momentum \cite{Sundaram1999}.
Phase space Berry curvatures are known to be important on the dynamics of
Bloch electrons, ubiquitously affecting equilibrium and transport properties
of solids \cite{Xiao2010,Nagaosa2010,Xiao2005,Sodemann2015,Moore2010,Xiao2012,Freimuth2013,Dong2020}.
It is therefore highly desirable to construct a semiclassical theory for
quasiparticle dynamics in superconductors, which systematically takes into
account these Berry curvatures, in order to provide an intuitive and effective
basis for analyzing various response properties of superconductors.

In this Letter, we introduce the semiclassical quasiparticle as a wave packet in the background of slowly varying
gauge potentials and the superconducting order parameter. Apart from Berry
curvatures inherited from the parent Bloch states, we identify new
contributions due to the non-conserving nature of the quasiparticle charge and
the phase space structure of the order parameter which is nontrivial in all
but the conventional s-wave superconductors. The quasiparticle also naturally
possesses a charge dipole moment, which can couple to a magnetic field through
the Lorentz force and induce field dependent Berry curvatures.

To demonstrate the utility of the semiclassical theory, we discuss how these
Berry curvatures modify the phase-space density of states of the
quasiparticles and the impact on electron tunneling spectroscopic
measurements. We also present results on thermal Hall conductivity due to
quasiparticles, and reveal its relationship with topological contribution from
the condensate that has been discussed extensively as a measure of topological
superconductors \cite{Senthil1999,read2000}. We assume spin conservation in this work for simplicity, and
illustrate our results using a twisted bilayer graphene model \cite{yuan2018} with a d+id
superconducting gap function for the order parameter.

{\it {\color{blue} Quasiparticle wave packet in second quantization formalism.}%
}---In order for a semiclassical theory of superconducting quasiparticles to be feasible, we assume
that all the possible inhomogeneities in the considered system are smooth in
the spread of a quasiparticle wave packet, whose center position is marked as
$\boldsymbol{r}_{c}$. For example, in the
mixed states of type-II superconductors, we focus only on the region far away
from the vortex core, where the pairing potential can be perceived as
slowly varying. A local Hamiltonian description of the wave packet hence emerges,
namely ${H}_{c}=\int d\boldsymbol{r}c_{\sigma\boldsymbol{r}}^{\dagger}(\hat{h}%
_{c}-\mu)c_{\sigma\boldsymbol{r}}-\int d\boldsymbol{r}d\boldsymbol{r}^{\prime}%
g(\boldsymbol{r}_{c}, \boldsymbol{r}-\boldsymbol{r}^{\prime})c_{\uparrow\boldsymbol{r}%
}^{\dagger}c_{\downarrow\boldsymbol{r}^{\prime}}^{\dagger}c_{\downarrow
\boldsymbol{r}^{\prime}}c_{\uparrow\boldsymbol{r}}$, where $c_{\sigma\boldsymbol{r}%
}^{\dagger}$ is the creation operator for an electron with spin $\sigma
(=\uparrow,\downarrow)$ at position $\boldsymbol{r}$, $\hat{h}_{c}\equiv
{h}_{0}\left(  \boldsymbol{r}\text{,}-i\nabla_{\boldsymbol{r}}-e\boldsymbol{A}%
(\boldsymbol{r}_{c},t);\{\beta_{i}(\boldsymbol{r}_{c})\}\right)  $ is the spin
degenerate single-electron Hamiltonian in the local approximation (set
$\hbar=1$), $\mu$ is the chemical potential, and $g$ is the effective
attractive interaction between electrons. We only consider spin-singlet
superconductors with intraband pairing and without spin-orbit coupling. The
slowly varying perturbation fields \{$\beta_{i}$\} ($i=1,2,..$) and the
electromagnetic vector potential $\boldsymbol{A}$ are represented by their values
at $\boldsymbol{r}_{c}$. $\hat{h}_{c}$
possesses local eigenfunctions $e^{ie\boldsymbol{A}(\boldsymbol{r}_{c},t)\cdot
\boldsymbol{r}}\psi_{n\sigma\boldsymbol{k};\boldsymbol{r_{c}}}(\boldsymbol{r})$, where
$\psi_{n\sigma\boldsymbol{k};\boldsymbol{r_{c}}}(\boldsymbol{r})$ are the local Bloch
functions of ${h}_{0}\left(  \boldsymbol{r}\text{,}-i\nabla_{\boldsymbol{r}%
};\{\beta_{i}(\boldsymbol{r}_{c})\}\right)  $, with local Bloch bands
$\xi_{n\boldsymbol{k;}\boldsymbol{r}_{c}}$. Here $n$ and $\boldsymbol{k}$ are the
indices for band (with two-fold spin degeneracy) and wave-vector,
respectively, and $\boldsymbol{r}_{c}$\ enters in the eigenstates parametrically
as a character of the local description.

The interaction term can be treated within a mean-field approach, ending with \cite{supp}
\begin{equation}
{H}_{c}=\sum_{n \sigma \boldsymbol{k}}  E_{n\boldsymbol{k};\boldsymbol{r}_{c}}%
\gamma_{n\sigma\boldsymbol{k};\boldsymbol{r}_{c}}^{\dagger}\gamma_{n\sigma
\boldsymbol{k};\boldsymbol{r}_{c}},
\end{equation}
where the creation/anihilation operators for the local eigenstate are
introduced by the Bogoliubov transformation, e.g., $\gamma_{n\uparrow
\boldsymbol{k};\boldsymbol{r}_{c}}^{\dagger}=\mu_{n\boldsymbol{k};\boldsymbol{r}_{c}}^{\ast
}c_{n\uparrow\boldsymbol{k};\boldsymbol{r}_{c}}^{\dagger}-\nu_{n\boldsymbol{k}%
;\boldsymbol{r}_{c}}^{\ast}c_{n\downarrow-\boldsymbol{k};\boldsymbol{r}_{c}}$. Here $c_{n\sigma\boldsymbol{k;}\boldsymbol{r}_{c}}^{\dagger}=\int d\boldsymbol{r}e^{ie\boldsymbol{A}%
(\boldsymbol{r}_{c},t)\cdot\boldsymbol{r}}\psi_{n\sigma\boldsymbol{k};\boldsymbol{r}_{c}%
}(\boldsymbol{r})c_{\sigma\boldsymbol{r}}^{\dagger}$
creates the local Bloch eigenstates of $\hat{h}_{c}$, whereas ($\mu
_{n\boldsymbol{k};\boldsymbol{r}_{c}}$, $\nu_{n\boldsymbol{k};\boldsymbol{r}_{c}}$)$^{T}$ and
$E_{n\boldsymbol{k};\boldsymbol{r}_{c}}=\sqrt{\xi_{n\boldsymbol{k};\boldsymbol{r}_{c}}%
^{2}+|\Delta_{n\boldsymbol{k};\boldsymbol{r}_{c}}|^{2}}$ are the Bogoliubov
wavefunction in this local Bloch representation and the eigenenergy,
respectively, and $\Delta_{n\boldsymbol{k};\boldsymbol{r}_{c}}$\ is the local
momentum-space superconducting pairing function. The quasiparticle operators not only define the excitations of the local Hamiltonian, but also determine the ground state of the local Hamiltonian with annihilation operators $|G\rangle= \mathcal{N} \prod_{n \sigma \boldsymbol{k}}  \gamma_{n\sigma
\boldsymbol{k};\boldsymbol{r}_{c}} |0\rangle$. Here $\mathcal{N}$ is the normalization factor and $|0\rangle$ is the vacuum for electrons.

Now we construct a quasiparticle
wave packet centered around ($\boldsymbol{r}_{c},\boldsymbol{k}_{c}$) with the local
creation operators acting on the superconducting ground state:
\begin{equation}
|\Psi_{n\uparrow}(\boldsymbol{r}_{c},\boldsymbol{k}_{c},t)\rangle=\int [d\boldsymbol{k}]
\alpha(\boldsymbol{k},t)\gamma_{n\uparrow\boldsymbol{k};\boldsymbol{r}_{c}}^{\dagger
}|G\rangle,
\label{wave-packet}
\end{equation}
where $\int [d\boldsymbol{k}]$ is shorthand for $\int d^m k /(2\pi)^m $ with $m$ the dimension of the system. The envelope function $\alpha(\boldsymbol{k},t)$ is sharply distributed in
reciprocal space so that it makes sense to speak of the wave-vector
$\boldsymbol{k}_{c}=\int [d\boldsymbol{k}]|\alpha(\boldsymbol{k},t)|^{2}\boldsymbol{k}$ of the
wave packet. We only demonstrate the spin-up
wave packet, as the spin-down case can be easily extrapolated.

{\it \color{blue} {Spin center and charge dipole of the wave packet.}}---For Bloch electrons the wave-packet center is simply the charge center.
However, superconducting quasiparticles are momentum-dependent
mixture of electrons and holes and thereby do not possess definite charges, rendering the charge center ill defined. On
the other hand, spin is a conserved quantity in the absence of spin-orbit
coupling, hence the spin center serves physically as the center of a
wave packet. For this purpose we consider the spin density operator ${\hat S} (\boldsymbol{r}) = c_{\uparrow, \boldsymbol{r}}^{\dagger}c_{\uparrow,\boldsymbol{r}}-c_{\downarrow
, \boldsymbol{r}}^{\dagger}c_{\downarrow,\boldsymbol{r}}$, and calculate its wave-packet averaging
$S(\boldsymbol{r}) = \langle \Psi|  {\hat S} (\boldsymbol{r}) |\Psi\rangle-\langle G|  {\hat S} (\boldsymbol{r}) |G\rangle$. This gives the distribution of spin on the wave packet, and its center, the spin center, is given by \cite{supp}
\begin{equation}
\boldsymbol{r}_{c} \equiv \int d\boldsymbol{ r}   S(\boldsymbol{ r}) \boldsymbol{ r}  = \frac{\partial\gamma_{c}}{\partial
\boldsymbol{k}_{c}}+\langle\phi|i\nabla_{\boldsymbol{k}_{c}}\phi\rangle-{\rho_{c}} \nabla_{\boldsymbol{k}_{c}}\theta_{c},
\end{equation}
where $\theta_{c}=\frac{1}{2}\arg\Delta_{n\boldsymbol{k}_{c};\boldsymbol{r}_{c}}$ is related to the phase of the
superconducting order parameter, $\rho_{c}=\xi_{n\boldsymbol{k}_c;\boldsymbol{r}_{c}%
}/E_{n\boldsymbol{k}_{c};\boldsymbol{r}_{c}}$ measures the non-conserved charge of the quasiparticles, $|\phi\rangle$ is the periodic part of the Bloch state
$|\psi_{n\sigma\boldsymbol{k}_{c};\boldsymbol{r}_{c}}\rangle$, and $\gamma_{c}=-\arg\alpha(\boldsymbol{k}_{c},t)$
is the phase of the envelope function. The Berry
connections contain not only the Bloch part $\mathcal{A}_{\boldsymbol{k}_{c}}%
^{b}=\langle\phi|i\nabla_{\boldsymbol{k}_{c}}\phi\rangle$ from the
single-electron band structure, but also the superconducting
part $\mathcal{A}_{\boldsymbol{k}_{c}}^{sc}=-{\rho_{c}} \nabla_{\boldsymbol{k}_{c}}\theta_{c}$ from the momentum dependence of the superconducting order parameter.

The spin center is not sufficient to describe the
coupling of quasiparticles with electromagnetic fields, which would inevitably involve information on the charge distribution upon the spread of a wave
packet. Since the charge distribution is not centered at $\boldsymbol{r}_{c}$,
there should be a charge dipole moment associated with a
wave packet. Indeed one can consider the charge density operator $\hat{Q} (\boldsymbol{r})= e(c_{\uparrow \boldsymbol{r}}^{\dagger}c_{\uparrow
\boldsymbol{r}}+c_{\downarrow \boldsymbol{r}}^{\dagger}c_{\downarrow \boldsymbol{r}})$, and its wave-packet averaging $Q(\boldsymbol{r}) = \langle \Psi|  {\hat Q} (\boldsymbol{r}) |\Psi\rangle-\langle G|  {\hat Q} (\boldsymbol{r}) |G\rangle$ provides a proper definition for the charge dipole moment \cite{supp}
\begin{equation}
\boldsymbol {d} \equiv  \int d \boldsymbol{ r} { Q} (\boldsymbol{r}) (\boldsymbol{ r} - {\boldsymbol r}_c)= e({\rho
_{c}^{2}-1})\frac{\partial\theta_{c}}{\partial\boldsymbol{k}_{c}}.
\end{equation}
It is nonzero only in the case of a momentum dependent phase of
superconducting order parameter. Furthermore, if the external-field-free
system has either time-reversal (space-inversion) symmetry, $\boldsymbol{d}$ is
an even (odd) function in momentum space, as can be inspected from the semiclassical
equations of motion proposed later.

{\it {\color{blue} Berry curvatures and semiclassical dynamics.}}--The
distinctive properties of the wave packet are anticipated to strongly
affect its semiclassical dynamics determined by the Lagrangian $\mathcal{L}%
=\langle\Psi|i\frac{d}{dt}-\hat{H}_{c}|\Psi\rangle - \langle G|i\frac{d}{dt}-\hat{H}_{c}|G \rangle$
\cite{Xiao2010}, and should be embodied in various Berry curvatures
characterizing the dynamical structure. Adopting the circular gauge
$\boldsymbol{A}(\boldsymbol{r}_{c})=\frac{1}{2}\boldsymbol{B}\times\boldsymbol{r}_{c}$,
which is suitable for the approximately uniform magnetic field in regions far
away from vortex lines, after some algebra we get \cite{supp} (hereafter the
wave packet center label $c$ is omitted for simplicity):
\begin{equation}
\mathcal{L}=-E + \boldsymbol{k}\cdot{\dot{\boldsymbol{r}}}  +(\mathcal{A}_{\boldsymbol{r}}^{b}- \rho\boldsymbol{v}^{s}+
\boldsymbol{B}\times\boldsymbol{\tilde{d}})\cdot{\dot{\boldsymbol{r}}}+(\mathcal{A}_{\boldsymbol{k}
}^{b}-{\rho}{\nabla_{\boldsymbol{k}}\theta})\cdot\dot{\boldsymbol{k}}.
\end{equation}
Here the coupling of the wave packet to the magnetic field
involves the charge dipole and gives $\boldsymbol{B}\times \boldsymbol{\tilde d}$, with $\boldsymbol{\tilde d} = \boldsymbol{{d}}/2$. Besides,
$\boldsymbol{v}^{s}=  {\nabla_{\boldsymbol{r}}\theta} -e\boldsymbol{A}$ is half of the
gauge invariant supercurrent velocity, and $\mathcal{A}_{\boldsymbol{r}}%
^{b}=\langle\phi|i\nabla_{\boldsymbol{r}}\phi\rangle$ is the real-space Berry
connection of the single-electron wave function.

The structure of the Lagrangian implies that the total Berry connections in
the momentum and real space take the forms of $\mathcal{A}_{\boldsymbol{k}%
}=\mathcal{A}_{\boldsymbol{k}}^{b}-{\rho}{\nabla_{\boldsymbol{k}}\theta}$ and
$\mathcal{A}_{\boldsymbol{r}}=\mathcal{A}_{\boldsymbol{r}}^{b}-\rho\boldsymbol{v}^{s}+\boldsymbol{B}\times\boldsymbol{\tilde d}$, respectively. Various Berry
curvatures are then formed as $\Omega_{\lambda_{\alpha}\lambda_{\beta}%
}=\partial_{\lambda_{\alpha}}\mathcal{A}_{\lambda_{\beta}}-\partial
_{\lambda_{\beta}}\mathcal{A}_{\lambda_{\alpha}}$, where $\boldsymbol{\lambda
}={\boldsymbol{r}}$, $\boldsymbol{k}$, and $\alpha$ and $\beta$ are Cartesian indices.
In particular, $\Omega_{k_{\alpha}k_{\beta}}$ and $\Omega_{r_{\alpha}r_{\beta
}}$\ are anti-symmetric tensors with respect to $\left(  \alpha\text{, }%
\beta\right)  $, whose vector forms read respectively
\begin{equation}
\boldsymbol{\Omega}_{\boldsymbol{k}}=i\langle\nabla_{\boldsymbol{k}}\phi|\times|\nabla_{\boldsymbol{k}}\phi\rangle-\nabla_{\boldsymbol{k}}\rho\times\nabla_{\boldsymbol{k}} \theta
\label{eq:berrycurvature}%
\end{equation}
and
\begin{equation}
\boldsymbol{\Omega}_{\boldsymbol{r}}=i\langle\nabla_{\boldsymbol{r}}\phi|\times|\nabla_{\boldsymbol{r}}\phi\rangle +e\rho\boldsymbol{B}- \nabla_{\boldsymbol{r}}\rho\times\boldsymbol{v}^{s}+\nabla_{\boldsymbol{r}}\times(\boldsymbol{B}\times\boldsymbol{\tilde d})%
.\label{eq:berrycurvature-realspace}
\end{equation}
One can readily verify that the above $\boldsymbol{\Omega}_{\boldsymbol{k}}$ coincides
with that obtained from the Bogoliubov-de Gennes equation \cite{liang2017wavepacket}.
The first terms in these two equations are the familiar Berry curvatures from the single-electron band structure \cite{Xiao2010}, while other terms involves superconductivity. Moreover, the characteristics of
superconductors, i.e., the charge non-conservation and the resultant charge
dipole of wave packet and the real-space supercurrent, are embedded in the last
three terms of $\boldsymbol{\Omega}_{\boldsymbol{r}}$.

Regarding the phase-space Berry curvature $\boldsymbol{\Omega}_{\boldsymbol{kr}}$, there are remarkable qualitative differences from that for Bloch
electrons, namely $\boldsymbol{\Omega}_{\boldsymbol{kr}}=0$ and $\boldsymbol{\Omega
}_{\boldsymbol{kr}}\neq0$ respectively in normal states and superconducting states
subjected to scalar perturbations. The underlying physics is that the scalar perturbation in the electronic
Hamiltonian is endowed with a spin structure in the Nambu space where the
quasiparticles live. Thus the usual scalar field felt by electrons is no longer
scalar for superconducting quasiparticles. Nonzero $\boldsymbol{\Omega}_{\boldsymbol{kr}}$ will play a vital
role in a number of experimental measurables \cite{Xiao2010}. For example, in the presence of pure magnetic perturbations, its trace reads
\begin{equation}\label{eq:Berrykr}
{\rm Tr} [\boldsymbol{\Omega}
_{\boldsymbol{kr}}] = -\nabla
_{\boldsymbol{k}}\rho\cdot \boldsymbol{v}^s -e\rho\boldsymbol{B} \cdot (\nabla_{\boldsymbol{k}}\rho\times\nabla_{\boldsymbol{k}} \theta).
\end{equation}
As will be shown later, this trace of the Berry-curvature tensor plays an important role in the geometric modulations to the quasiparticle local density of states \cite{Xiao2010}.

With the above Berry curvatures, the Euler-Lagrange
equations of motion for superconducting quasiparticles possess the same noncanonical structure as for
Bloch electrons \cite{Sundaram1999,Xiao2010}. Having realized this, we neglect the Berry curvatures from Bloch band structures for simplicity and focus on those originated from superconductivity. Thus the equations of motion read:
\begin{widetext}
\begin{eqnarray}\label{eq:semiclassical}
&&{\dot {\boldsymbol r}}=\nabla_{\boldsymbol k} E +  \dot{\boldsymbol k}  \times ( \nabla_{\boldsymbol{k}}\rho\times\nabla_{\boldsymbol{k}}\theta ) +  \nabla_{\boldsymbol k}    (\rho{\boldsymbol v}^{ s}-  {\boldsymbol B} \times {\boldsymbol {\tilde d}} ) \cdot \dot {\boldsymbol  r}  -{ \dot{\boldsymbol r}} \cdot\nabla_{{\boldsymbol r}} ({\rho} {\nabla_{\boldsymbol k} \theta} ),
\\\nonumber
&&{\dot {\boldsymbol k}}=-\nabla_{\boldsymbol r} E
+{\dot {\boldsymbol r}} \times (e\rho\boldsymbol{B}-\nabla_{\boldsymbol{r}}\rho\times\boldsymbol{v}^{s}+\nabla_{\boldsymbol{r}}\times(\boldsymbol{B}\times\boldsymbol{\tilde d}))
- \nabla_{{\boldsymbol r}} ({\rho}   {\nabla_{\boldsymbol k} \theta}  ) \cdot {{ \dot{ \boldsymbol k}}}+ { \dot{\boldsymbol k}} \cdot\nabla_{{\boldsymbol k}} (  \rho{\boldsymbol v}^{ s}-  {\boldsymbol B} \times {\boldsymbol {\tilde d}}  )
.
\end{eqnarray}
\end{widetext}
In the absence of superconductivity, $\rho = 1$, $\boldsymbol{\tilde d } =0$, and $\theta=0$, hence the equations of motion reduce to the usual ones for electrons \cite{Sundaram1999}.
It is also worthwhile to mention that, for trivial superconducting pairing, the momentum-space Berry connection vanishes but the real-space one may still survive due to the
supercurrent velocity: $\mathcal{A}_{\boldsymbol{r}}=-\rho\boldsymbol{v}^{s}$. The
resulting Berry curvature in real space is given by
$\boldsymbol{\Omega}_{\boldsymbol{r}}=e\rho\boldsymbol{B}+\nabla_{\boldsymbol{r}}\rho
\times\boldsymbol{v}^{s}$. The equations of motion
describe the quasiparticle dynamics subjected to background super-flow, and take a similar form to those for bosonic Bogoliubov quasiparticles in a Bose-Einstein
condensate with a vortex \cite{zhang2006BdGberry}.

Equation (\ref{eq:semiclassical}) is the central result of this work. It provides a framework to understand quasiparticle dynamics in superconductors subjected to various perturbations. In the following, we apply this semiclassical theory to calculate several properties of superconductors.

{\it {\color{blue} Density of states.}}--A most direct consequence of the Berry curvatures appearing in the equations of motion is the breakdown of the phase-space volume conservation. As a result, the phase-space measure $\mathcal{D}(\boldsymbol{r,k})$ is modified by
Berry curvatures \cite{Xiao2005}, which to the first order of the spatial inhomogeneity can be expressed as
\begin{eqnarray}
\mathcal{D}(\boldsymbol{r,k}) = 1 + {\rm Tr} {\boldsymbol \Omega}_{\boldsymbol kr} - {\boldsymbol \Omega}_{\boldsymbol r} \cdot {\boldsymbol \Omega}_{\boldsymbol k}.
\end{eqnarray}
The modification may originate from various perturbations, such as the supercurrent and magnetic field. We note that $\partial{\mathcal{D}}/\partial{\boldsymbol B}=0$ since the relevant terms in ${\rm Tr} {\boldsymbol \Omega}_{\boldsymbol kr}$ and ${\boldsymbol \Omega}_{\boldsymbol r} \cdot {\boldsymbol \Omega}_{\boldsymbol k}$ cancel each other, in sharp contrast to the case of Bloch electrons \cite{Xiao2005}.

$\mathcal{D}$ would influence the quasiparticle local density of states $n({\boldsymbol r}, \omega)$, which is just the integration of the phase-space volume with the fixed quasiparticle energy $E$,
\begin{equation}
n({\boldsymbol r}, \omega) = \int [d{\boldsymbol k}] \mathcal{D}(\boldsymbol{r,k}) (|\mu|^2 \delta(\omega - E_{{\boldsymbol r},{\boldsymbol k}} )+ |\nu|^2 \delta(\omega + E_{{\boldsymbol r},{\boldsymbol k}})).
\end{equation}
This quasiparticle density of states is proportional to the differential conductance which
can be directly measured by scanning tunneling microscopy \cite{fischer2007}.
For instance, in the case of a small supercurrent, we have $\mathcal{D} = 1  -\nabla
_{\boldsymbol{k}}\rho\cdot \boldsymbol{v}^s $ according to Eq. (\ref{eq:Berrykr}), which gives the modulation part as
\begin{equation}
\delta n(  {\boldsymbol r},\omega) =  -   \int [d{\boldsymbol k}]\boldsymbol{v}^s \cdot  \nabla
_{\boldsymbol{k}}\rho    (|\mu|^2 \delta(\omega - E_{{\boldsymbol r},{\boldsymbol k}} )+ |\nu|^2 \delta(\omega + E_{{\boldsymbol r},{\boldsymbol k}})) ,
\end{equation}
where $\delta n = n - n_0 $ with $n_0 =  \int [d{\boldsymbol k}] ((|\mu|^2 \delta(\omega - E_{{\boldsymbol r},{\boldsymbol k}} )+ |\nu|^2 \delta(\omega + E_{{\boldsymbol r},{\boldsymbol k}}))$ being the local density of states given by the conventional formula. This modification to the density of states depends on the direction of the supercurrent, hence could be experimentally verified by injecting supercurrent on different directions.

{\it {\color{blue} Thermal Hall transport.}}-- The semiclassical theory can
also be employed to study the transport properties in superconductors such
as the intrinsic thermal Hall effect. Compared to the Green's function
method \cite{Qin2011,sumiyoshi2013}, the semiclassical theory has an
advantage of subtracting conveniently the circulating magnetization current
\cite{Cooper1997} without a detailed calculation of the energy magnetization
\cite{Xiao2020EM}. Here we sketch the key steps from the semiclassical
equations towards the thermal Hall transport. We start from the
semiclassical expression for the local energy current density $\boldsymbol{j}%
^{\text{Q}}=\int [d{\boldsymbol{k}}]\mathcal{D}({\boldsymbol{k}})f\left( E_{{%
\boldsymbol{k}}},T\right) E_{{\boldsymbol{k}}}\dot{\boldsymbol{r}}$ \cite%
{local} where $f\left( E_{{\boldsymbol{k}}},T\right) $ is the Fermi-Dirac
distribution at temperature $T$. Then we substitute the equation of motion
for $\dot{\boldsymbol{r}}$ \cite{note-derivation}, and find $\boldsymbol{j}^{%
\text{Q}}=-{\nabla }T\times \frac{\partial }{\partial T}\int [d{\boldsymbol{k%
}]}h\boldsymbol{\Omega }_{\boldsymbol{k}}+{\nabla }\times \int [d{%
\boldsymbol{k}}]h\boldsymbol{\Omega }_{{\boldsymbol{k}}}$ where we introduce
the auxiliary function $h\left( E_{{\boldsymbol{k}}},T\right) =-\int_{E_{{%
\boldsymbol{k}}}}^{\infty }d\eta f\left( \eta ,T\right) \eta $. Now the
second term is a circulating current which should be discounted, leaving the
transport current $\boldsymbol{j}^{\text{Q,tr}}=\int [d{{\boldsymbol{k}}}]%
\frac{\partial h}{\partial T}\boldsymbol{\Omega }_{\boldsymbol{k}}\times {%
\nabla }T$. The Hall response of this current is given by
\begin{equation}\label{eq:kappa}
\kappa^{\text{Q}}
_{xy}=\frac{2}{T}\int [d \boldsymbol{k}]  (\boldsymbol{\Omega}_{\boldsymbol{k}})_{z} \int_{ E_{\boldsymbol k}}^\infty  d\eta\eta^{2}f'\left(  \eta,T\right) ,
\end{equation}
where the factor 2 denotes the spin degeneracy.

The above formula only accounts for the contribution from quasiparticles
beyond the superconducting condensate. It is physically reasonable to make
the connection $\kappa _{0}+\kappa^{\text{Q}}_{xy}=\kappa _{xy}^{BdG}$ between this
\textquotedblleft quasiparticle plus condensate\textquotedblright\
description and the Bogoliubov-de Gennes (BdG) one \cite{sumiyoshi2013},
Here $\kappa _{0}$ is the thermal Hall conductivity contributed by the
condensate and
$\kappa _{xy}^{BdG}=\frac{1}{T}\int [d \boldsymbol{k}]  (\boldsymbol{\Omega}_{\boldsymbol{k}})_{z} \left(\int_{ E_{\boldsymbol k}}^\infty - \int_{ -E_{\boldsymbol k}}^\infty\right)  d\eta\eta^{2}f'\left(  \eta,T\right)$
is the conductivity obtained
using the particle-hole symmetric BdG bands. In $\kappa _{xy}^{BdG}$ the
spin degeneracy and the particle-hole redundancy cancel out, and $-E_{%
\boldsymbol{k}}$ means the BdG   ''valence
band''  whose Berry curvature is $-(\boldsymbol{\Omega }_{\boldsymbol{k}})_{z}$. Therefore, the condensate contribution reads
\begin{equation}
\kappa _ 0=-\frac{1}{T}\int [d \boldsymbol{k}]  (\boldsymbol{\Omega}_{\boldsymbol{k}})_{z} \int_{ -\infty}^\infty  d\eta\eta^{2}f'\left(  \eta,T\right)=\frac{\pi C_{1}k_{B}^{2}T}{6\hbar},
\end{equation}
where the summation over
momentum gives exactly the Chern number $C_{1}$. This recovers the quantized thermal Hall
conductance given by edge-state analysis \cite{Senthil1999,read2000}. Having
clarified the above relationship, in the following we use the simplified
notation $\kappa _{0}+\kappa^{\text{Q}}_{xy}\rightarrow \kappa _{xy}$ to represent the
total thermal Hall conductivity.

{\it {\color{blue} Model illustration: twisted-bilayer graphene with d + id
superconductivity.}}-- To illustrate the application of the semiclassical theory, we consider
the twisted-bilayer graphene system which has been proposed to support a topological chiral d-wave superconducting state \cite{balents2020,guo2018d+id,liu2018d+id,chen2020d+id}.
We take the effective four-band tight-binding Hamiltonian to describe the system \cite{yuan2018},
\begin{eqnarray}
\hat{H}_{0}  &&  =-\mu\sum_{i}\tilde{c}_{i}^{\dagger}\tilde{c}_{i}+t_{1}%
\sum_{\langle i,j\rangle}\tilde{c}_{i}^{\dagger}\tilde{c}_{j}+t_{2}%
\sum_{[i,j]}\tilde{c}_{i}^{\dagger}\tilde{c}_{j}\nonumber\\
&&  +t{}_{3}\sum_{[i,j]}\tilde{c}_{i}^{\dagger}[i\sigma_{y}\otimes\sigma
_{0}]\tilde{c}_{j}+h.c. \label{model}%
\end{eqnarray}
where $\tilde{c}_{i}^{\dagger}\equiv(c_{i,x,\uparrow}^{\dagger}%
,c_{i,y,\uparrow}^{\dagger},c_{i,x,\downarrow}^{\dagger},c_{i,y,\downarrow
}^{\dagger})$ is the electron creation operator with two distinct orbitals
$\alpha=(p_{x},p_{y})$, $\sigma_{y}$ is the Pauli matrix, $\sigma_{0}$ is the
identity matrix, $t_{i}$ ($i=1,2,3$) are hopping parameters, and $\langle
i,j\rangle$ and $[i,j]$ represent the summations over the three nearest
neighbor lattice vectors and over the second-nearest neighbors within the same
sublattice, respectively. We diagonalize this Hamiltonian and take the band with the dispersion function
$
\xi(\boldsymbol{k}%
)=-|t_{1}h_{1}(\boldsymbol{k})|+2t_{2}h_{2}(\boldsymbol{k}) +2t{}_{3}%
h_{3}(\boldsymbol{k})-\mu,
$
where $h_{1}(\mathbf{k})=1+2e^{i\frac{3}{2}k_{x}}\cos(\frac
{\sqrt{3}}{2}k_{y})$ is from the nearest-neighbor hopping, $h_{2}(\mathbf{k}%
)=\cos(3k_{x})+\cos(-\frac{3}{2}k_{x}+\frac{3\sqrt{3}}{2}k_{y})+\cos(-\frac
{3}{2}k_{x}-\frac{3\sqrt{3}}{2}k_{y})$ and $h_{3}(\mathbf{k})=\sin
(3k_{x})+\sin(-\frac{3}{2}k_{x}+\frac{3\sqrt{3}}{2}k_{y})+\sin(-\frac{3}%
{2}k_{x}-\frac{3\sqrt{3}}{2}k_{y})$ are from the next nearest-neighbor hopping.
Superconductivity in twisted bilayer graphene with $d_{x^{2}-y^{2}}+id_{xy}$ pairing symmetry can be described
by the superconducting gap function in the form of \cite{jiang2008d+id,nandkishore2012,Black_Schaffer2014}
$
\Delta(\boldsymbol{k})=\sum_{i=1}^{3}\Delta_{i}%
\cos(\boldsymbol{k}\cdot\boldsymbol{R}_{i}-\varphi_{\boldsymbol{k}}),
$
where
$(\Delta_1,\Delta_2,\Delta_3) \equiv \sqrt{\frac{1}{6}}(2 \Delta,-\Delta+i\sqrt{3}\Delta^{\prime},-\Delta-i\sqrt{3}\Delta^{\prime})$ with
$\Delta$ and $\Delta ^{\prime}$ being the superconducting gap amplitudes for $d_{x^{2}-y^{2}}$ and $d_{xy}$ pairing, respectively, $\varphi_{\boldsymbol{k}}=\arg[h_{1}(\boldsymbol{k)]}$ is the phase of the nearest-neighbor hopping, and
$\boldsymbol{R}_{i}$ are the three nearest-neighbor lattice vectors.

\begin{figure}[htb]
\begin{center}
\includegraphics[clip = true, width =\columnwidth]{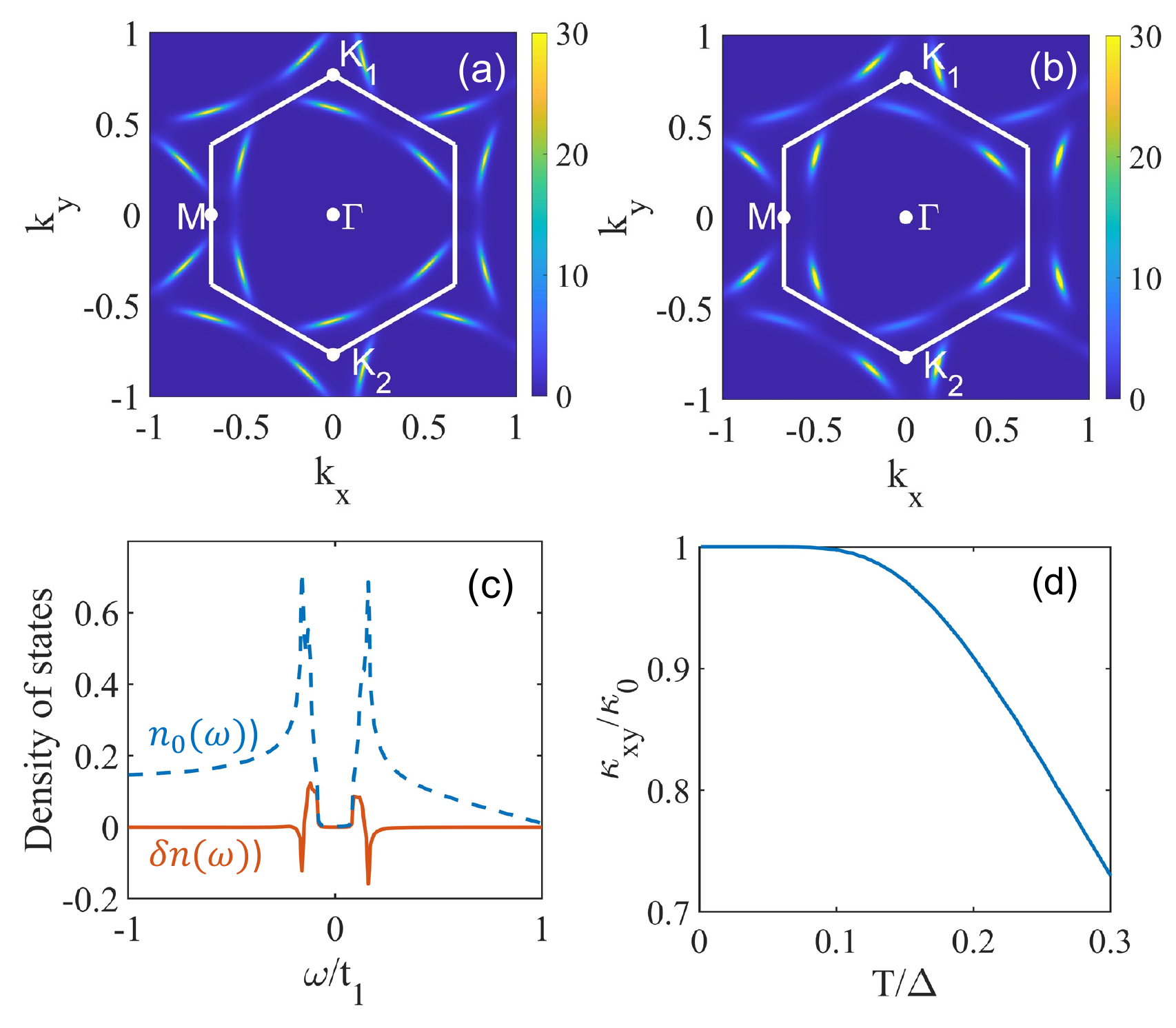}
\end{center}
\caption{Berry curvatures of the tight-binding model for twisted-bilayer graphene with (a) symmetric
$d_{x^{2}-y^{2}}$ and $id_{xy}$ superconducting gaps
$\Delta= \Delta^{\prime}$ and (b) asymmetric superconducting gaps
$\Delta= 2\Delta^{\prime}$. Model parameters are taken as $t_{2}/t_{1} = 0.05$, $t_{3}/t_{1} = 0.2$, $\mu=-0.9 t_{1}$ and $\Delta/t_{1}= 0.1$.
(c) Berry curvature modification to the quasiparticle density of states $\delta n(\omega)$ (solid line) with a constant supercurrent of $\nabla_{\boldsymbol r} \theta=\frac{\sqrt{2}\pi}{10}\hat x$. The conventional density of states $n_0(\omega)$ is demonstrated for comparison (dashed line). (d) The thermal Hall conductivity as a function of temperature. Parameters for (c) and (d) are taken the same as those for (a).}
\label{fig:berrycurvature}
\end{figure}

Now we can calculate the momentum-space Berry curvature by Eq. (\ref{eq:berrycurvature}) for this tight-binding model. In Fig.~\ref{fig:berrycurvature}a, we demonstrate the Berry curvature with typical band parameters given in Ref. \cite{yuan2018} and symmetric $d_{x^2-y^2}$ and $d_{xy}$ gaps.
The band structure of the tight-binding model has trivial topology, and the Berry curvatures are entirely contributed by the
superconducting gap function. Because of the particle-hole symmetry in
superconductors, the Berry curvatures concentrate around the Fermi
surface. This is clearly shown in Fig.~\ref{fig:berrycurvature}a., where the
Berry curvature has symmetric peaks reflecting the $D_{3}$ symmetry of the lattice structure and the
gap function. In Fig.~\ref{fig:berrycurvature}b we show the result with
asymmetric $d$ and $id$ pairing gaps. For this case the superconducting gap breaks the rotational symmetry, leaving only the reflectional symmetry with respect to the $k_x$-axis for the Berry curvature distribution.

As a simple example, we study the density of states modulation due to a supercurrent which could originate from the injected current or a superconducting vortex. As shown in Fig.~\ref{fig:berrycurvature}c, the obtained modulation $\delta n$ is quite considerable in comparison with the non-perturbed density of states $n_0$. We also note that this modulation depends on both the amplitudes and the direction of the supercurrent, and can have much richer pattern if other perturbations are introduced.

We also calculate the intrinsic thermal Hall transport of the toy model,
demonstrating the temperature dependence of the thermal Hall conductivity, with $\kappa_0$ as its zero temperature value. As shown in Fig.~\ref{fig:berrycurvature}d, the ratio of the thermal Hall conductivity to $\kappa_0$ has a near exponential dependence on the temperature at the low temperature regime, and becomes an approximated linear function at higher temperatures. These features would be helpful for identifying the $d+id$ paring in twisted-bilayer graphene systems.

Finally, we note that the results shown in Fig.~\ref{fig:berrycurvature} are obtained with a single band, while for the tight-binding model there are two bands intersecting with the chemical potential. The thermal conductance from the two bands are exactly the same, while the modulations to the local density of states have a sign reversal and a resultant cancellation. In order to observe the modulation to the local density of states in twisted-bilayer graphene system, band or momentum resolved tunneling experiments are required.

In summary, we derived the semiclassical equations of motion for superconducting quasiparticle wave packets, and identified various Berry curvature contributions in momentum space, real space as well as phase space. We demonstrated the power of the theory with examples such as the density of states modulation and the thermal Hall transport, and applied the theory to study the twisted-bilayer graphene system. Our theory opens up a new route to study rich Berry-phase effects on equilibrium and
transport properties of superconducting quasiparticles.

\textit{Acknowledgments.---} We thank Zhongbo Yan, Tianxing Ma, Huaiming Guo, and Jihang Zhu for very valuable discussions. This work was supported by NKRDPC-2017YFA0206203, 2017YFA0303302, 2018YFA0305603,
NSFC (Grant No. 11774435), and Guangdong Basic and Applied Basic Research Foundation (Grant No. 2019A1515011620). The work at The University of Texas at Austin was supported by NSF (EFMA-1641101) and Robert A. Welch Foundation (F-1255).

\bibliography{Feynman}

\begin{thebibliography}{10}
\bibitem{note-derivation}%
  \BibitemOpen
  \bibinfo {howpublished} {First we have
  $\boldsymbol{j}^{\text{Q}}=\int_{\boldsymbol{k}}\mathcal{D}\frac{\partial
  h}{\partial E}\dot {\boldsymbol{r}}$. Plugging the equations of motion and
  the phase-space measure up to the first order of spatial gradient, and
  integrating by parts, we get the result.}%
  \bibAnnoteFile{Stop}{note-derivation}%
\end{thebibliography}

\end{document}